\documentclass[12pt]{article}

\usepackage{epsf}
\usepackage{amsmath}
\usepackage[dvips]{graphicx}

\begin{document}

\begin{center}
{\bfseries DEUTERON POLARIZATION OBSERVABLES IN ELASTIC $\vec{d}p$ SCATTERING AT INTERMEDIATE
ENERGIES}

\vskip 5mm

M.A. Shikhalev$^{\dag}$

\vskip 5mm

{\small {\it Veksler and Baldin Laboratory of High Energies, JINR, Dubna}
\\
$\dag$ {\it E-mail: shehalev@theor.jinr.ru}}
\end{center}

\vskip 5mm

\begin{center}
\begin{minipage}{150mm}
\centerline{\bf Abstract} A calculation of vector and tensor polarization observables of a
deuteron in an optical potential framework for elastic proton-deuteron scattering at the
incident deuteron energy $E_d=880$ MeV is presented. The main aim is to point out the role of
various effects in explaining the wide-range angle distribution of these observables. Under the
investigation are a double-scattering mechanism in the optical potential, the influence of
various $NN$-partial waves and as well a comparison of calculations with two different deuteron
wave-functions derived from the Bonn-CD $NN$-potential model and dressed bag model of
Moscow-Tuebingen group.
\end{minipage}
\end{center}

\vskip 10mm

\section{Introduction}
The reaction of elastic nucleon-deuteron scattering are considered both by experimentalists and
theoreticians as one of the clue problems in few-nucleon physics. For three decades it has been
served as a hope to obtain more information about intermediate- and short-range $NN$ interaction
and as a probe of the deuteron structure at small distances (for a review, see Ref.
\cite{Gloekle}). During the last decade it is also studied with the purpose of testing of
various $3N$-forces and particularly their spin-dependance \cite{Sekiguchi,Uzikov}. It is also
of interest as a basic reaction to establish a polarimetry for vector-tensor mixed polarized
beams.

Now, there is a set of high quality nucleon-nucleon potentials that describe two-nucleon elastic
scattering data perfectly below the pion production threshold. But they underestimate the three
nucleon bound energy. The incorporation of a $3N$ force in the calculations improves the
situation but some unsolved questions remained, especially it concerns its spin-dependant
properties. The study of polarization observables in elastic $pd$ scattering, as expected, can
provide an additional information about the properties of $3N$ force. To describe $pd$ elastic
scattering below the pion production threshold different techniques have been applied. Among
them the momentum space Faddeev equations can now be solved with high accuracy for the most
modern two- and three-nucleon forces. The $3N$ forces in such a calculations are of
Fujita-Miyazawa \cite{Fujita,Urbana} or Tucson-Melbourne \cite{Coon} type. Also, in Ref.
\cite{Mach} a coupled-channel model with excitation of the stable $\Delta$-isobar was
successfully applied to $pd$ scattering employing Chebyshev expansion of two-body amplitudes.

But calculations at incident proton energies greater than $200$ MeV in lab encounter with some
nontrivial difficulties. The first problem is that up to now there is no reliable quantitative
model for $NN$-interaction above the inelastic threshold. All existing models that pretend to
description of two-nucleon scattering up to 1 GeV do it only in semi-quantitative  manner
\cite{Mach2}. The second problem is that it is no longer conventional three-body problem above
the pion production threshold. The nucleon isobar and meson degrees of freedom start to play a
significant role. And last but not least, there are purely computational difficulties in
performing relativistic three-body calculations at higher energies \cite{Gloeckle2}. Hence one
must resort to some kind of simplifications and approximations in the study of this reaction.

In this work, the optical potential framework to calculate polarization observables of a
deuteron is used. The incident deuteron energy is taken to be 880 MeV in lab. This energy is
just below the region where the influence of $\Delta$-isobar excitation becomes significant
masking a large momentum component of the deuteron wave function \cite{Uzikov2}. Moreover, this
energy lies in the interval of energies proposed for the measurements of $dp$ elastic scattering
analyzing powers at internal target station of the Nuclotron at JINR \cite{Ladygin}.
\section{Theoretical framework}
The basic equations for three-body problem in quantum physics are the Faddeev equations which
can be written in operator form:
\begin{equation}
\label{Fadd}
 U=PG_0^{-1}+PTG_0U,
\end{equation}
where $U=U_{\mu_d'\mu_N',\mu_d\mu_N}(\vec{q'},\vec{q})$ is an amplitude of elastic
$dp$-scattering, $\mu_d,\mu_N$ are spin quantum numbers, $T$ -- $NN$ scattering matrix,
$G_0=(E-H_0+i\epsilon)^{-1}$ is a free propagator of $3N$ system and $P\equiv
P_{12}P_{23}+P_{13}P_{23}$ stands for a permutation operator that takes into account the
property of identity of three nucleons. Here and further in this work we have dropped out all
$3N$ force contributions.

One can rewrite (\ref{Fadd}) in the form that is more appropriate for calculations at higher
energies where an impulse approximation is justified:
\begin{equation}
\label{Optic} U=V_{\rm opt}+V_{\rm opt}G_dU.
\end{equation}
This is the optical potential framework and $V_{\rm opt}$ is a nucleon-deuteron optical
potential:
\begin{equation}
\label{potent}
 V_{\rm opt}=PG_0^{-1}+PT_cG_0V_{\rm
opt}.
\end{equation}
Here $T_c$ is a $NN$ $T$-matrix with subtracted deuteron pole term and $G_d$ is a deuteron
contribution in the spectral decomposition of the two-body Hamiltonian.

\begin{figure}[h]
\begin{center}
\includegraphics[width=15cm,height=7cm]{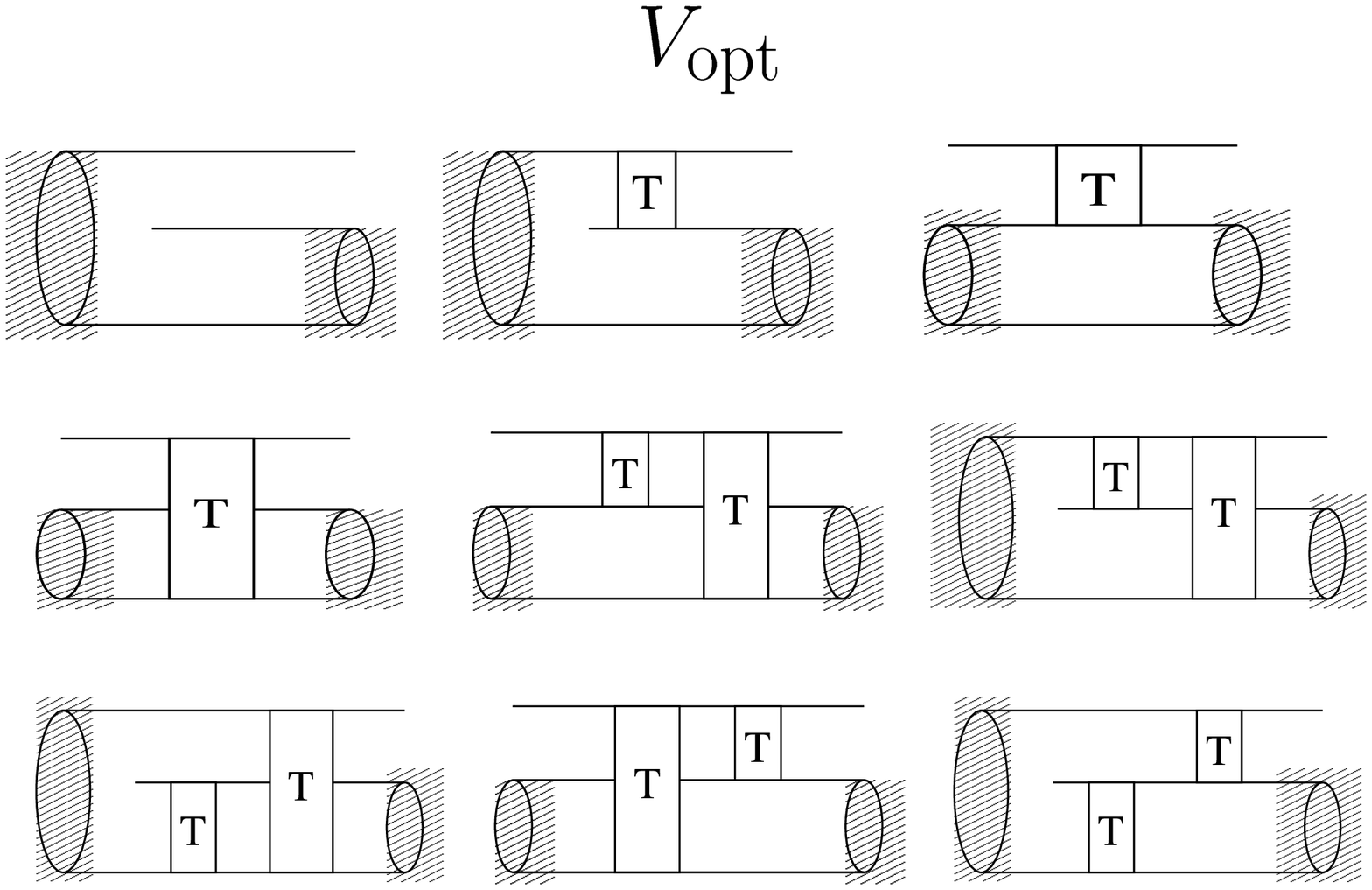}
\caption{Pictorial representation of $pd$ optical potential up to the second order in the series
expansion of Eq. (\ref{potent}).\label{fig}}
\end{center}
\end{figure}

Although, the Eq. (\ref{Optic}) seems quite simple and represents a common two-body scattering
equation, the derivation of the optical potential by solving Eq. (\ref{potent}) is as difficult
as solving Faddeev equations themselves. However, as was mentioned in Introduction, at energies
higher than the pion production threshold, new degrees of freedom, other than nucleons, come
into the game. But this new physics is not taken into account in (\ref{potent}). Also, these are
the energies at which the impulse approximation is a good starting point, so one can make a
series expansion of $V_{\rm opt}$ in (\ref{potent}), keeping a possibility to include in $V_{\rm
opt}$ some additional terms:
$$V_{\rm opt}=PG_0^{-1}+PT_cP+PT_cG_0PT_cP+\ldots$$
Pictorially it can be described as the sum of nine graphs (up to the second order in series
expansion), see Fig. \ref{fig}.
The first diagram is a usual one-nucleon exchange mechanism, the next three diagrams are so
called triangle diagrams and represent the single-scattering process and the last five diagrams
take into account the inelastic break up channel.

The one-nucleon exchange contribution is written in a usual way:
\begin{equation*}
{}_{2(31)}\langle
f|G_0^{-1}|i\rangle=-\left[\sqrt{q^2+M_d^2}-\sqrt{q^2+M_N^2}-\sqrt{4q^2\cos^2\frac{\theta}{2}+M_N^2}\
\right]\Psi_{13}\left(\vec{q}+\!\!\!\!\fontsize{15}{\baselineskip}\begin{array}{c}\frac{1}{2}\end{array}\!\!\!\vec{q'}\right)\Psi_{23}
\left(\vec{q'}+\!\!\!\!\fontsize{15}{\baselineskip}\begin{array}{c}\frac{1}{2}\end{array}\!\!\!\vec{q}\right).
\end{equation*}
Here, we omit a complicated notation related to a summation over spin and orbital quantum
numbers. $\Psi_{23}$ is a wave function of the deuteron composed of nucleons 2 and 3.

To evaluate a single scattering diagram, one must implement an integration over the internal
momentum of nucleons in the deuteron. To do this, a knowledge about the fully off-shell behavior
of $NN$ $T$-matrix is required:
\begin{equation*}
{}_{1(23)}\langle f|T_c|i\rangle=\int\frac{\rm{d}^3p}{(2\pi)^3}
\Psi_{23}\left(\vec{p}+\!\!\!\!\fontsize{15}{\baselineskip}\begin{array}{c}\frac{1}{4}\end{array}\!\!\!\vec{k}\right)
T_c\left(\vec{q'},\vec{p}-\!\!\!\!\fontsize{15}{\baselineskip}\begin{array}{c}\frac{3}{4}\end{array}\!\!\!\vec{q'}+
\!\!\!\!\fontsize{15}{\baselineskip}\begin{array}{c}\frac{1}{4}\end{array}\!\!\!\vec{q};
\vec{q},\vec{p}-\!\!\!\!\fontsize{15}{\baselineskip}\begin{array}{c}\frac{3}{4}\end{array}\!\!\!\vec{q}+
\!\!\!\!\fontsize{15}{\baselineskip}\begin{array}{c}\frac{1}{4}\end{array}\!\!\!\vec{q'}\right)
\Psi_{23}\left(\vec{p}-\!\!\!\!\fontsize{15}{\baselineskip}\begin{array}{c}\frac{1}{4}\end{array}\!\!\!\vec{k}\right),
\end{equation*}
here $k=\vec{q}-\vec{q'}$ is a transferred momentum.

However, the absence of high quality $NN$ interaction model above the inelastic threshold at the
present moment forces anyone to make some approximate evaluation of the integral, proceeding
from the assumption of either an off-shell behavior of somehow parameterized $T$-matrix or
taking only its on-shell value. In this work, the $T$-matrix is put on-shell in the way that
minimizes the off-shell corrections. This is an optimal impulse approximation
\cite{Gurvitz,Neil}:
$$T_c\left(\vec{q'},\vec{p}-\!\!\!\!\fontsize{15}{\baselineskip}\begin{array}{c}\frac{3}{4}\end{array}\!\!\!\vec{q'}+
\!\!\!\!\fontsize{15}{\baselineskip}\begin{array}{c}\frac{1}{4}\end{array}\!\!\!\vec{q};
\vec{q},\vec{p}-\!\!\!\!\fontsize{15}{\baselineskip}\begin{array}{c}\frac{3}{4}\end{array}\!\!\!\vec{q}+
\!\!\!\!\fontsize{15}{\baselineskip}\begin{array}{c}\frac{1}{4}\end{array}\!\!\!\vec{q'}\right)\longrightarrow
T_c\left(\vec{q'},-\!\!\!\fontsize{15}{\baselineskip}\begin{array}{c}\frac{3}{4}\end{array}\!\!\!\vec{q'}+
\!\!\!\!\fontsize{15}{\baselineskip}\begin{array}{c}\frac{1}{4}\end{array}\!\!\!\vec{q};
\vec{q},-\!\!\!\fontsize{15}{\baselineskip}\begin{array}{c}\frac{3}{4}\end{array}\!\!\!\vec{q}+
\!\!\!\!\fontsize{15}{\baselineskip}\begin{array}{c}\frac{1}{4}\end{array}\!\!\!\vec{q'};E_{eff}\right).$$
Then $T$-matrix is evaluated in the so called $NN$ Breit frame, where an effective scattering
energy is $$E_{\rm{eff}}=\frac{s}{2M_N}-2M_N,$$
$$s=\left(\sqrt{q^2+M_N^2}+\sqrt{\frac{q^2}{8}(5-3\cos\theta)+M_N^2}\right)^2-\frac{q^2}{4}\cos^2\frac{\theta}{2}.$$
For example, at deuteron laboratory energy $T_{\rm{lab}}=880$ MeV and scattering angle
$\theta=10^\circ$ the effective scattering energy
 of two nucleons is $E_{\rm{eff}}=443$ MeV, whereas at the angle $\theta=170^\circ$ -- $E_{\rm{eff}}=796$
 MeV. The numerical values of the $T$-matrix are obtained from partial wave analysis SAID \cite{Arndt} that
 provides the $NN$ phase shifts up to 3 GeV and the total angular momentum
 $J_{NN}=7$.

This permits the $T$-matrix to be shifted outside the integral and the remaining integration
produces the deuteron form factor:
$${}_{1(23)}\langle
f|T_c|i\rangle=T_c\left(\vec{q'},-\!\!\!\fontsize{15}{\baselineskip}\begin{array}{c}\frac{3}{4}\end{array}\!\!\!\vec{q'}+
\!\!\!\!\fontsize{15}{\baselineskip}\begin{array}{c}\frac{1}{4}\end{array}\!\!\!\vec{q};
\vec{q},-\!\!\!\fontsize{15}{\baselineskip}\begin{array}{c}\frac{3}{4}\end{array}\!\!\!\vec{q}+
\!\!\!\!\fontsize{15}{\baselineskip}\begin{array}{c}\frac{1}{4}\end{array}\!\!\!\vec{q'};E_{eff}\right)S(\vec{k}).$$

For the double scattering term one has two integrals -- one over the internal momentum in the
deuteron and another over the intermediate momentum of the scattered nucleon:
\begin{multline}
{}_{1(23)}\langle
f|T_2G_0T_1|i\rangle=\int\frac{\rm{d}^3p}{(2\pi)^3}\frac{\rm{d}^3p'}{(2\pi)^3}\frac{\rm{d}^3q''}{(2\pi)^3}
\delta^3\left(\vec{p'}+\vec{p}-\!\!\!\!\fontsize{15}{\baselineskip}\begin{array}{c}\frac{1}{2}\end{array}\!\!\!\vec{q''}+
\!\!\!\!\fontsize{15}{\baselineskip}\begin{array}{c}\frac{1}{4}\end{array}\!\!\!\vec{q'}+
\!\!\!\!\fontsize{15}{\baselineskip}\begin{array}{c}\frac{1}{4}\end{array}\!\!\!\vec{q}\right)\nonumber\\
\times\Psi_{23}\left(\!\!\!\!\fontsize{15}{\baselineskip}\begin{array}{c}\frac{\vec{p'}-\vec{p}}{2}\end{array}\!\!\!-
\!\!\!\!\fontsize{15}{\baselineskip}\begin{array}{c}\frac{3}{8}\end{array}\!\!\!\vec{q'}-
\!\!\!\!\fontsize{15}{\baselineskip}\begin{array}{c}\frac{1}{8}\end{array}\!\!\!\vec{q}+
\!\!\!\!\fontsize{15}{\baselineskip}\begin{array}{c}\frac{1}{2}\end{array}\!\!\!\vec{q''}\right)
T_2\left(\vec{q'},\vec{p'}-\!\!\!\!\fontsize{15}{\baselineskip}\begin{array}{c}\frac{3}{4}\end{array}\!\!\!\vec{q'}+
\!\!\!\!\fontsize{15}{\baselineskip}\begin{array}{c}\frac{1}{4}\end{array}\!\!\!\vec{q''};
\vec{q''},\vec{p'}-\!\!\!\!\fontsize{15}{\baselineskip}\begin{array}{c}\frac{3}{4}\end{array}\!\!\!\vec{q''}+
\!\!\!\!\fontsize{15}{\baselineskip}\begin{array}{c}\frac{1}{4}\end{array}\!\!\!\vec{q'}\right)G_0 \label{double}\\
\times
T_1\left(\vec{q''},\vec{p}-\!\!\!\!\fontsize{15}{\baselineskip}\begin{array}{c}\frac{3}{4}\end{array}\!\!\!\vec{q''}+
\!\!\!\!\fontsize{15}{\baselineskip}\begin{array}{c}\frac{1}{4}\end{array}\!\!\!\vec{q};
\vec{q},\vec{p}-\!\!\!\!\fontsize{15}{\baselineskip}\begin{array}{c}\frac{3}{4}\end{array}\!\!\!\vec{q}+
\!\!\!\!\fontsize{15}{\baselineskip}\begin{array}{c}\frac{1}{4}\end{array}\!\!\!\vec{q''}\right)
\Psi_{23}\left(\!\!\!\!\fontsize{15}{\baselineskip}\begin{array}{c}\frac{\vec{p'}-\vec{p}}{2}\end{array}\!\!\!+
\!\!\!\!\fontsize{15}{\baselineskip}\begin{array}{c}\frac{3}{8}\end{array}\!\!\!\vec{q}+
\!\!\!\!\fontsize{15}{\baselineskip}\begin{array}{c}\frac{1}{8}\end{array}\!\!\!\vec{q'}-
\!\!\!\!\fontsize{15}{\baselineskip}\begin{array}{c}\frac{1}{2}\end{array}\!\!\!\vec{q''}\right),\nonumber
\end{multline}

Taking into account in $G_0$ only its pole part and employing the impulse approximation
\begin{align}
&T_2\left(\vec{q'},\vec{p'}-\!\!\!\!\fontsize{15}{\baselineskip}\begin{array}{c}\frac{3}{4}\end{array}\!\!\!\vec{q'}+
\!\!\!\!\fontsize{15}{\baselineskip}\begin{array}{c}\frac{1}{4}\end{array}\!\!\!\vec{q''};
\vec{q''},\vec{p'}-\!\!\!\!\fontsize{15}{\baselineskip}\begin{array}{c}\frac{3}{4}\end{array}\!\!\!\vec{q''}+
\!\!\!\!\fontsize{15}{\baselineskip}\begin{array}{c}\frac{1}{4}\end{array}\!\!\!\vec{q'}\right)G_0
T_1\left(\vec{q''},\vec{p}-\!\!\!\!\fontsize{15}{\baselineskip}\begin{array}{c}\frac{3}{4}\end{array}\!\!\!\vec{q''}+
\!\!\!\!\fontsize{15}{\baselineskip}\begin{array}{c}\frac{1}{4}\end{array}\!\!\!\vec{q};
\vec{q},\vec{p}-\!\!\!\!\fontsize{15}{\baselineskip}\begin{array}{c}\frac{3}{4}\end{array}\!\!\!\vec{q}+
\!\!\!\!\fontsize{15}{\baselineskip}\begin{array}{c}\frac{1}{4}\end{array}\!\!\!\vec{q''}\right)\nonumber\longrightarrow\\
&-i\pi\frac{2M_N}{3q}\delta(q''-q)T_2\left(\vec{q'},-\!\!\!\fontsize{15}{\baselineskip}\begin{array}{c}\frac{3}{4}\end{array}\!\!\!\vec{q'}+
\!\!\!\!\fontsize{15}{\baselineskip}\begin{array}{c}\frac{1}{4}\end{array}\!\!\!\vec{q''};
\vec{q''},-\!\!\!\fontsize{15}{\baselineskip}\begin{array}{c}\frac{3}{4}\end{array}\!\!\!\vec{q''}+
\!\!\!\!\fontsize{15}{\baselineskip}\begin{array}{c}\frac{1}{4}\end{array}\!\!\!\vec{q'};E_{eff2}\right)
T_1\left(\vec{q''},-\!\!\!\fontsize{15}{\baselineskip}\begin{array}{c}\frac{3}{4}\end{array}\!\!\!\vec{q''}+
\!\!\!\!\fontsize{15}{\baselineskip}\begin{array}{c}\frac{1}{4}\end{array}\!\!\!\vec{q};
\vec{q},-\!\!\!\fontsize{15}{\baselineskip}\begin{array}{c}\frac{3}{4}\end{array}\!\!\!\vec{q}+
\!\!\!\!\fontsize{15}{\baselineskip}\begin{array}{c}\frac{1}{4}\end{array}\!\!\!\vec{q''};E_{eff1}\right),\nonumber
\end{align}
finally one gets
\begin{multline}{}_{1(23)}\langle f|T_2G_0T_1|i\rangle=-\frac{2i}{3}\pi
qM_N\int\frac{\rm{d}^2\Omega_{q''}}{(2\pi)^3}T_2\left(\vec{q'},-\!\!\!\fontsize{15}{\baselineskip}\begin{array}{c}\frac{3}{4}\end{array}\!\!\!\vec{q'}+
\!\!\!\!\fontsize{15}{\baselineskip}\begin{array}{c}\frac{1}{4}\end{array}\!\!\!\vec{q''};
\vec{q''},-\!\!\!\fontsize{15}{\baselineskip}\begin{array}{c}\frac{3}{4}\end{array}\!\!\!\vec{q''}+
\!\!\!\!\fontsize{15}{\baselineskip}\begin{array}{c}\frac{1}{4}\end{array}\!\!\!\vec{q'};E_{eff2}\right)\nonumber\\
T_1\left(\vec{q''},-\!\!\!\fontsize{15}{\baselineskip}\begin{array}{c}\frac{3}{4}\end{array}\!\!\!\vec{q''}+
\!\!\!\!\fontsize{15}{\baselineskip}\begin{array}{c}\frac{1}{4}\end{array}\!\!\!\vec{q};
\vec{q},-\!\!\!\fontsize{15}{\baselineskip}\begin{array}{c}\frac{3}{4}\end{array}\!\!\!\vec{q}+
\!\!\!\!\fontsize{15}{\baselineskip}\begin{array}{c}\frac{1}{4}\end{array}\!\!\!\vec{q''};E_{eff1}\right)S\left(\vec{q''}-
\frac{1}{2}(\vec{q'}+\vec{q})\right).\nonumber
\end{multline}

The scattering equation (\ref{Optic}) is solved in helicity basis employing $K$-matrix
approximation, i.e. only the pole part of the two-body propagator $G_d$ is remained thus all
terms in the equation contain only on-shell information about the optical potential and
scattering amplitude:
\begin{multline}
\langle\lambda'_d\lambda'_N|U^J(q',q)|\lambda_d\lambda_N\rangle=
\langle\lambda'_d\lambda'_N|V_{\mathrm{opt}}^J(q',q)|\lambda_d\lambda_N\rangle-
i\frac{A(q,q)}{q}\langle\lambda'_d\lambda'_N|V_{\mathrm{opt}}^J(q',q)U^J(q,q)|\lambda_d\lambda_N\rangle+\nonumber\\
\frac{2}{\pi}{\cal
P}\int\frac{\rm{d}q''}{q^2-q''^2}\biggl(\langle\lambda'_d\lambda'_N|V_{\mathrm{opt}}^J(q',q'')U^J(q'',q)|\lambda_d\lambda_N\rangle
A(q'',q) -
\langle\lambda'_d\lambda'_N|V_{\mathrm{opt}}^J(q',q)U^J(q,q)|\lambda_d\lambda_N\rangle
A(q,q)\biggr),\nonumber
\end{multline}
where the kinematical factor is
$$A(q'',q)=q''^2\frac{(E_q+E_{q''})((E_q^2+E_{q''}^2)/2-q^2-q''^2-M_N^2-M_d^2)}{E_q^2+E_{q''}^2}.$$

Then for the scattering amplitude in spin space one has (the incident particle is going along
the $z$-axis):
\begin{multline}
\langle\mu'_d\mu'_N|U(q,\theta)|\mu'_d\mu'_N\rangle=4\pi\sum_J\sum_{\lambda'_d,\lambda'_N}(-1)^{\lambda'_d-\mu_d}
(2J+1)d^{1/2\ast}_{\mu'_N,\lambda'_N}(\theta)d^{1\ast}_{\mu'_d,-\lambda'_d}(\theta)\nonumber\\
\times
d^{J}_{\lambda'_N-\lambda'_d,\mu_N+\mu_d}(\theta)\langle\lambda'_d\lambda'_N|U^J(q,q)|-\mu_d\mu_N\rangle.\nonumber
\end{multline}
It was found that at $T_{\mathrm{lab}}=880$ MeV a convergent result is obtained at $J=37/2$.
\section{Results}
\begin{figure}[b]
\includegraphics[width=\textwidth]{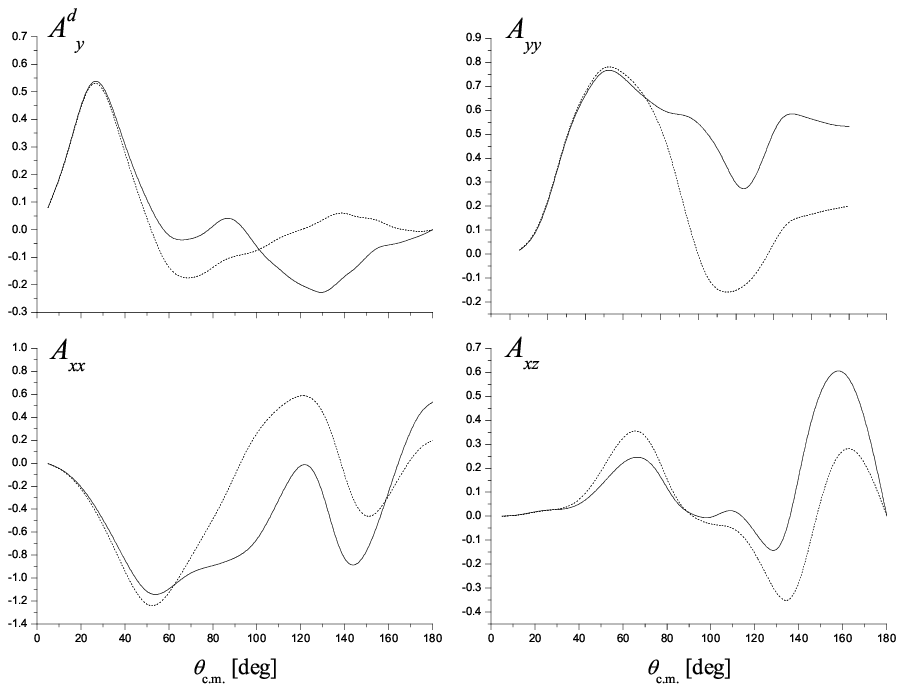}
\caption{Deuteron vector and tensor polarization observables at the energy $E_d=880$ MeV in lab.
The solid and dashed curves are calculations with the deuteron wave function derived in the
Bonn-CD model and DB model of $NN$ interaction respectively.\label{fig1}}
\end{figure}

The calculation of the polarization observables is performed with to different deuteron wave
functions. The first one is a wave function derived in the meson-exchange Bonn-CD model
\cite{Machleidt}. The general trait of wave functions of this kind, derived from the most modern
$NN$ potentials, is their depletion at small distances. The other possible choice is the wave
function with a nodal behavior. This node corresponds to a so-called forbidden state in $NN$
system as a consequence of the six-quark dynamics and the fact that the mostly symmetric
six-quark state $|s^6\rangle$ has a small $NN$ component \cite{Obukh}. As one of the
representatives of such a wave functions with a nodal behavior can serve the wave function of
dressed bag (DB) model of Moscow-Tuebingen group \cite{Kukulin}. In Fig. \ref{fig1} a comparison
of two calculations with these different wave function is presented. As one can see, the
remarkable difference in the polarization observables occurs at large scattering angles at which
the high momentum component of the wave functions is probed. It should be noted however, that
even if we can discriminate this two choices of the wave functions by experimental data, to
obtain an ultimate conclusion, which of the models are preferable, one has to include in
calculations a corresponding $3N$ force. And the $3N$ forces are different in these two
concepts.

\begin{figure}[h]
\includegraphics[width=\textwidth]{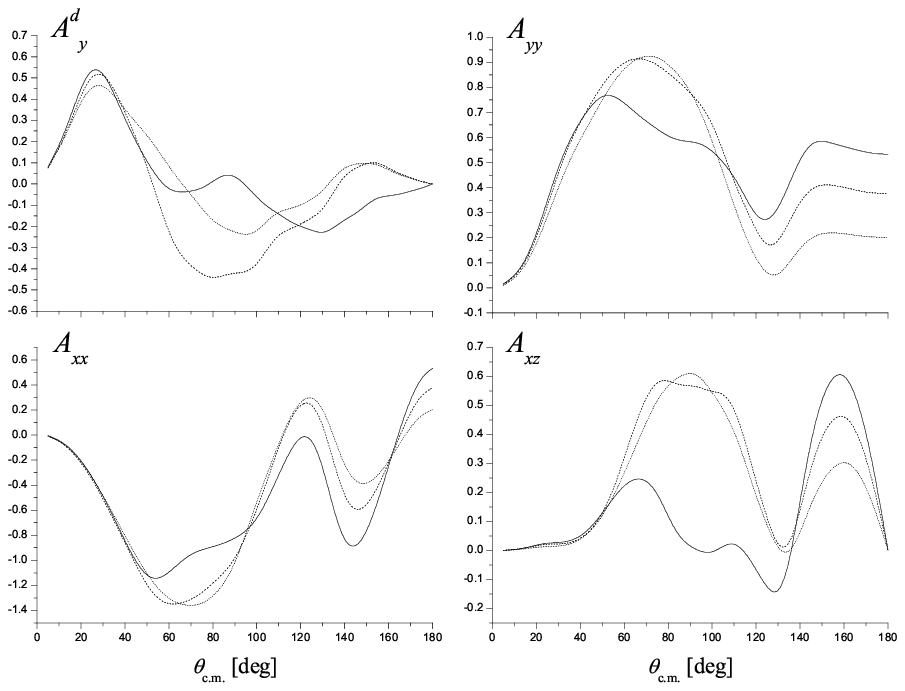}
\caption{The effect of a break up channel in the optical potential - dashed curve, and when all
rescattering mechanisms are omitted - dotted curve. The solid curve is a full calculation with
Bonn-CD deuteron wave function.\label{fig2}}
\end{figure}

In Fig. \ref{fig2} the influence of rescattering and break up channel on the polarization
observables at the deuteron energy 880 MeV is shown. It is evident that taking into account the
break up process or double-scattering in the optical potential is inevitable to obtain
quantitative predictions at these energies. The rescattering terms which are imbedded in the
$pd$ scattering equation is not so pronounced and much less important, however not negligible.

\begin{figure}[h]
\includegraphics[width=\textwidth]{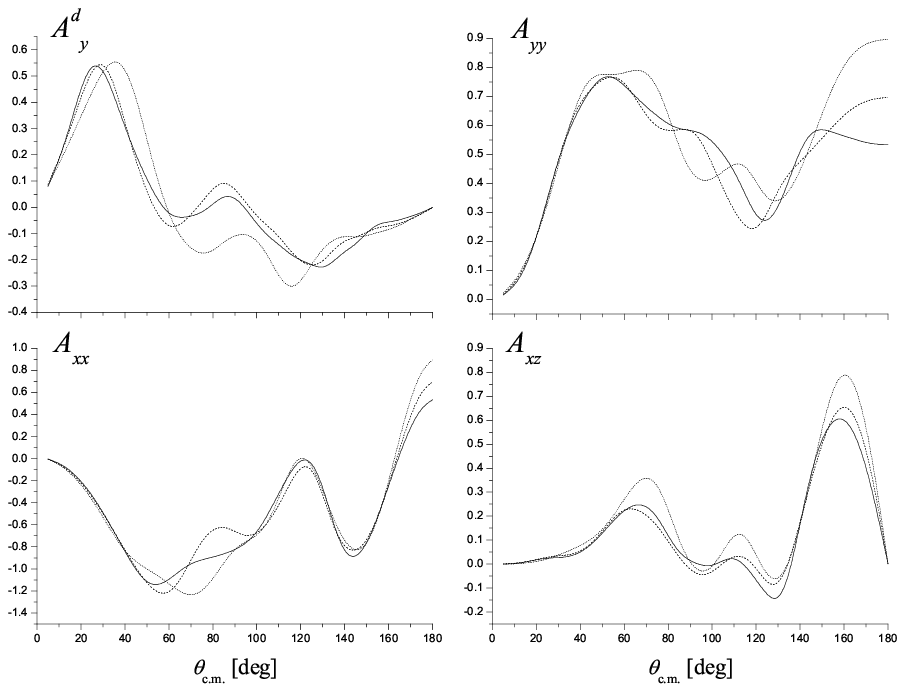}
\caption{The results of calculations that take into account different amount of partial waves in
$NN$ $T$-matrix. The solid curve is a result when all $J_{NN}=7$ SAID partial waves are included
in calculation, whereas dashed and dotted curves are the calculations with only $J_{NN}=5$ and
$J_{NN}=3$ respectively.\label{fig3}}
\end{figure}

Finally, the contribution from different partial waves in $NN$ $T$-matrix is examined, see
Fig.~\ref{fig3}. The full calculation takes into account all $NN$ partial waves that
parameterized in SAID partial wave analysis \cite{Arndt}, i.e. $J_{NN}=7$. The other two
calculations are done with total angular momentum up to $J_{NN}=5$ - dashed curve and $J_{NN}=3$
- dotted curve. Although at some scattering angles the differences between the observables are
very small, the convergence is only partial. But it seems that $J_{NN}=7$ is sufficient angular
momentum at this energy range.

\section{Summary and Conclusions}
A calculation of the deuteron polarization observables $A^d_y$, $A_{yy}$, $A_{xx}$ and $A_{xz}$
through an optical potential model for elastic proton-deuteron scattering at the incident
deuteron energy $E_d=880$ MeV was presented. Under investigation were the double-scattering
mechanism in the optical potential, the effect from precise treatment of two-body unitarity in
proton-deuteron Lippman--Schwinger equation, the influence of various $NN$-partial waves and as
well a comparison of calculations with two different deuteron wave-functions derived from the
Bonn-CD $NN$-potential model \cite{Machleidt} and DB model of Moscow-Tuebingen group
\cite{Kukulin}. For the $NN$ input, the model independent approach in which nucleon-nucleon
$T$-matrix is taken to be on-shell and evaluated in the so-called Breit frame, was used. The
calculation was carried out in the helicity partial wave decomposition and total angular momenta
up to $J=37/2$ in proton-deuteron system seem to be sufficient to obtain a convergent result. As
for $NN$-partial waves, we took all SAID partial waves up to $J_{NN}=7$. The large effect from
double-scattering mechanism at 880 MeV as well as the large dependency on the short-range
behavior of the deuteron wave function were observed. \vspace{0.5cm}

This work is partly supported by the Russian Foundation for Basic Research, grant 04-02-17107a.

\end{document}